\documentclass[preprint]{aastex}

\shorttitle{Spectral Line Gradients in NGC 3610}
\shortauthors{Howell, Brodie, Strader, Forbes, \& Proctor}

\begin{document}


\title{Probing Spectral Line Gradients Beyond One Effective Radius in NGC 3610}


\author{Justin H. Howell}
\affil{Lick Observatory, University of California,
 Santa Cruz, CA 95064, USA}
\email{jhhowell@ucolick.org}

\author{Jean P. Brodie}
\affil{Lick Observatory, University of California,
 Santa Cruz, CA 95064, USA}
\email{brodie@ucolick.org}

\author{Jay Strader}
\affil{Lick Observatory, University of California,
 Santa Cruz, CA 95064, USA}
\email{strader@ucolick.org}

\author{Duncan A. Forbes}
\affil{Centre for Astrophysics \& Supercomputing, Swinburne University,
  Hawthorn, VIC 3122, Australia}
\email{dforbes@astro.swin.edu.au}

\author{Robert Proctor}
\affil{Centre for Astrophysics \& Supercomputing, Swinburne University,
  Hawthorn, VIC 3122, Australia}
\email{rproctor@astro.swin.edu.au}


\begin{abstract}

The outer region ($0.75$--$1.25~r_e$ in the $B$-band) of the 
merger-remnant elliptical
NGC~3610 is studied using extremely high signal to noise Keck spectra, with
a supplementary spectrum of the galaxy center.  Stellar
population parameters --- age, [Z/H], $[\alpha/{\rm Fe}]$ --- are
measured in several apertures along the slit.  Using the multi-index
simultaneous fitting method of Proctor et~al. (2004), no significant
stellar population gradients are detected in the outer parts of the galaxy.  
The overall gradients relative to the galaxy center are consistent with 
those found in many other early-type galaxies, though the metallicity 
gradient is much steeper than would be expected if NGC~3610 formed in a 
major merger event.  Standard analysis methods using the H$\beta$ index 
are found to produce spurious radially variable gradients.

\end{abstract}


\section{Introduction}

Many studies of the star formation histories of elliptical galaxies are
restricted to central apertures \citep[e.~g.][]{jorg97, kd98, tfwg1, tf02}.
The necessary observations are relatively simple, and allow useful
measurements of the basic physical parameters (velocity dispersion, age, 
metallicity, and $\alpha$-enhancement) in the centers of galaxies.
However, radial gradients in these quantities provide dramatically
greater leverage when comparing to galaxy formation models.  The 
observational challenges are correspondingly great, particularly for
gradients extending beyond one effective radius ($r_e$).
Several studies to date \citep[e.~g.][]{carollo93, davies93, ka99, mehlert03}
have surveyed gradients in different samples of galaxies.  All are in
agreement that most ellipticals become less metal-rich with radius by
approximately $-0.2$~dex per decade in $r/r_e$.  As this gradient is
much shallower than that predicted by monolithic collapse galaxy formation
models \citep{carlberg} the standard interpretation is that merger events
dilute the gradients of monolithically-forming precursor galaxies
\citep{white80}.  The simulations of a large number of elliptical
galaxies under CDM initial conditions by \citet{kobayashi04} confirm
these conclusions using more sophisticated and up to date physical
models.  None of the galaxies with the steepest metallicity gradients, 
$<-0.35$~dex per decade in radius, had undergone a major merger, and the
average gradient for merger remnant galaxies was shallower than the
average gradient for galaxies which formed via monolithic collapse.
The simulations also indicated that the extent to which the metallicity
gradient will change in a merger event depends on both the mass ratio
of the two merging galaxies and the gas fraction in the two galaxies.

This study uses extremely high signal-to-noise ($S/N$) spectra to study 
gradients in the stellar population parameters of NGC~3610 out to
$1.25~r_e$.  NGC~3610 is an interesting and nearby (distance modulus 
$m-M=32.65$; Tonry et~al. 2001) elliptical galaxy.
Because of the extraordinary amount of fine structure \citep[7.6][]{ss92}, the
most in any galaxy outside of the Toomre sequence \citep{toomre72}, 
NGC~3610 has long been suspected of being an intermediate--age merger remnant.
The anomalously blue $B-V$ color of the galaxy \citep{goud94, idiart02}
suggests recent star formation.  The photometric study of \citet{silbot98}
confirmed the presence of intermediate--age stars.  Although Silva \& Bothun
argued that the cold stellar disk lying along the major axis \citep{sb90, 
wmsf97, rw92} is difficult to reconcile with predictions of a major merger 
origin for NGC~3610, the simulations of \citet{barnes02} have 
shown that such disks are a natural by-product of mergers of gas-rich systems.

Spectral line index measurements of the center of NGC~3610 \citep{hg04, 
denicolo}
indicate that it contains a young and metal-rich stellar population. 
Furthermore, young and old globular cluster populations have been found
in this system \citep{wmsf97, strader03, strader04}.  The young
globular cluster population is of comparable age to the galaxy center,
but of slightly lower metallicity.  Many of the old globular clusters are
metal-rich, indicating that at least one progenitor galaxy must have had a
significant bulge population.

If NGC~3610 is a remnant of a recent major merger event, it should have a
metallicity gradient even shallower than that found in other early-type
galaxies.  The simulations of \citet{bekki} predict gradients of between
$-0.03$ and $-0.13$~dex per decade in radius in galaxies formed from mergers of
late-type (bulgeless) spiral galaxies.  No simulations are yet available
for metallicity gradients in mergers involving early-type spirals.

The paper is organized as follows.  In \S~2 we describe the data and
reduction procedures, along with detailed data analysis.  Stellar
population parameters are derived in \S~3, and
\S~4 discusses the conclusions of this work.

\section{Data}

\subsection{Observation and Data Reduction}

Data were obtained using LRIS-B \citep{lris} on the Keck telescope in 2003 
January and 2004 April as by-products of an ongoing program to measure the 
ages and metallicities of the globular cluster systems of many nearby 
elliptical galaxies, including NGC~3610.  The April data includes the center 
of NGC~3610 with the $162''$ long slit lying
along the major axis.  The January slit was oriented at an intermediate angle
between NGC 3610's major and minor axes and was off-center by $10''$.  
Figure~1 shows the slit orientation relative to the galaxy for this
observation.  This slit position was chosen to intersect two globular cluster 
candidates.  In both cases the galaxy was positioned in the center of one 
half of the CCD, such that
the galaxy signal would be read out entirely by one of the two amplifiers.
Five $1800$~s exposures were taken resulting in a total exposure time of 
2.5~hours for the off-axis (January) data; the on-axis (April) data consist 
of a single 5~minute exposure.  Seeing was $\sim0.8''$ for the off-axis data
and $\sim0.6''$ for the on-axis spectrum.  Spectral resolution is 
approximately $70$~km/s, with a wavelength range from 
$3000\mbox{\AA}$--$5500\mbox{\AA}$, a dispersion of $0.63\mbox{\AA}$/pixel, 
and a spatial scale of $0.135''$/pixel.  Flat fields, flux standards, and Lick 
index standard stars were also observed throughout each observing run.

Data reduction followed standard procedures.  First,
the LRIS bias subtraction routine was executed.
Flat-fielding was performed for wavelengths greater
than $\sim 3700\mbox{\AA}$ only, as the flat-field lamp produces almost no
light at bluer wavelengths.  All data bluer than this flat-field cutoff
were ignored in subsequent analysis.  Next, the sections
of the CCD read out by each amplifier were corrected to a common dispersion
and intensity scale.
Wavelength calibration was performed in two dimensions using the standard
longslit packages {\sc identify, reidentify, fitcoords}, and {\sc transform}
in IRAF\footnote{IRAF is distributed by the National Optical Astronomy 
Observatories, which are operated by the Association of Universities for 
Research in Astronomy, Inc., under cooperative agreement with the 
National Science Foundation.}.  Similar procedures to identify arc lamp
reference lines and calculate the dispersion solution between pixel space
and wavelength space exist in any other astronomical data reduction package.
Cosmic rays were masked on the {\sc transform}ed images. 
After {\sc transform} the galaxy spectrum was almost perfectly aligned with 
the CCD.  The on-axis data was extracted, traced, and sky subtracted using 
the {\sc apall} task.  The broad spatial brightness profile of the off-axis 
data precluded sufficiently accurate centroid measurement for an aperture to 
be traced by {\sc apall}, so extraction was performed by combining rows of 
the CCD directly.  A sky aperture was defined consisting of a $13.5''$
region at the far edge of the CCD from the galaxy spectrum.
At over $7 r_e$ the contribution of
galaxy light to this sky spectrum is completely negligible.  This sky spectrum 
was subtracted from each row of each of the galaxy apertures
before the galaxy spectra were extracted.  

Six galaxy apertures were defined 
with $S/N$ per pixel of 120 in each.  At such high $S/N$, Poisson errors
have become negligible compared to the various sources of calibration error,
not to mention the internal uncertainties within the Lick/IDS system.
The minimum aperture size was several times as large as the seeing.
The outermost apertures were restricted to go no further than the point at 
which the galaxy intensity fell below twice the sky intensity.  These
apertures have been labeled in pairs.  The innermost pair is `A', the
middle pair is `B', and the outer pair is `C'.  Within each pair, the 
aperture closer to the minor axis of the galaxy is `1' while the aperture
closest to the major axis is `2'.  For the on-axis spectrum, a central 
aperture was defined extending to $r_e/8$, and two flanking apertures
between $r_e/8$ and the above sky cutoff ($\sim 1~r_e$ in this case)
were also extracted using the trace from the $r_e/8$ aperture.  All of these 
extractions were averages set to exclude 
the cosmic rays and bad pixels which were masked earlier.
After extraction, the one-dimensional spectra were flux calibrated using 
standard techniques.  

\subsection{Velocity Dispersion Correction and Lick Index Measurements}

When measuring spectral indices in galaxies, it is necessary to correct
for the effect of the galaxy's velocity dispersion broadening the absorption
lines.  This is critical both for meaningful comparisons to stellar population
models and for comparisons between different apertures, which in general
have slightly different velocity dispersions.
This correction was done by matching model spectra from \citet{ricardo}
to the galaxy spectra.  The major difference between these model spectra
and previous work is the use of the Jones spectral library \citep{jones}
for deriving the fitting functions for spectra of particular ages and
metallicities.  Unlike other models based on this library, the atmospheric
parameters for the library stars are entirely new.  In addition, these 
models provide synthetic SSP spectra, which are used in the velocity 
dispersion correction.  The velocity dispersion ($\sigma$) for each aperture 
was measured using the Pixfit code \citep{pixfit}, which broadens a
model spectrum by a user-defined range of $\sigma$ values and determines
the best fit by chi-squared minimization between the broadened model
spectra and the data.  
The resulting $\sigma$ measurements were used to smooth the best fitting models 
in order to derive a $\sigma$-correction.  The galaxy and model
spectra (both for $\sigma = 0$ and $\sigma = \sigma_{3610}$ varying from 
$135$ to $180$~km/s depending on aperture) were degraded
to match the Lick/IDS resolution \citep{wo97}.  The smoothing for the
blue part of the spectrum was performed piecewise in $100\mbox{\AA}$ increments,
as the Lick/IDS resolution varies strongly with wavelength in that region.

Equivalent widths of the Lick/IDS indices defined in \citet{trager98} and 
\citet{wo97} were measured using a modified 
version of the {\tt bwid} program provided by R.~M.~Rich \citep{bwid}.
A sample spectrum (not degraded to IDS resolution) is shown in Figure~2 
with the Lick/IDS indices and the location of [OIII]$\lambda{5007}$ marked.
For each line index, the multiplicative correction applied to the galaxy 
measurements was the ratio of the $\sigma = 0$ model index to the model 
index smoothed to the galaxy's velocity dispersion.  
Lick index standard stars (six from the off-axis observing run, three from
the on-axis observing run) were reduced identically to the galaxy spectra,
and offsets were determined to calibrate these data to the Lick system.
The RMS scatter of the standard star index measurements relative to the
Lick system is $0.03$--$0.08\mbox{\AA}$ for most indices, with Fe5015 and
H$\delta_{\rm F}$ being more uncertain.
The H$\delta_{\rm F}$ calibration is especially uncertain due to a poor match 
of standard stars to the galaxy in that index.

NGC~3610 shows very little emission at [OIII]$\lambda{5007}$.  However, in a 
few apertures from the off-axis spectra this emission, while very small, is 
non-zero and may affect several absorption lines.
The [OIII] line was used to correct H$\beta$ via Equation~2 of 
\citet{tfwg1}.  The H$\delta_{\rm F}$, H$\gamma_{\rm F}$, and Fe5015 indices
were corrected as described in \citet{kuntschner}.  This measurement was 
done independently for each of the five exposures; the mean value and the 
error in the mean for each aperture were used in all emission correction 
calculations.

The index measurements for each aperture are presented in Table~1 along
with the total observational error calculated below.  Also listed are the 
measurements from the combined spectrum of all six off-axis apertures
(Fig.~2).  The luminosity-weighted median radius of 
each aperture is presented in Table~2 in units of the effective
radius at that angle from the center of the galaxy.  The mean $B$-band 
effective radius of NGC~3610 was taken from \citet{faber89} while the axis 
ratio was measured from a direct image taken along with the off-axis spectra.

\subsection{Errors}

The total errors for each index in each aperture were measured as follows.
The five independently reduced observations provide a direct measure of
the random Poisson error for the off-axis apertures.  Sky subtraction error 
was measured by determining the index offsets when the sky level was over-
or undersubtracted by 1.4\%, the estimated Poisson uncertainty in the sky
intensity.  Wavelength calibration error contributes
significantly in this instrument and lamp setup.  The best wavelength
solution had an rms error of $0.40\mbox{\AA}$.  This was translated into index 
errors by shifting the spectra by that amount and noting the extent of the 
offset.  Radial velocity uncertainty was slightly less than wavelength 
calibration error in magnitude, and was treated identically.  Velocity 
dispersion corrections are known to be very index-dependent \citep{trager98}.  
The fitting functions from that work were used to translate a conservative 
estimate ${\sigma}_{\sigma} \approx 10$~km/s into index errors.  Finally, the 
error in measuring [OIII]$\lambda{5007}$ emission was estimated as described
above and propagated into $\sigma_{{\rm H}\beta}$, $\sigma_{{\rm H}\gamma}$,
$\sigma_{{\rm H}\delta}$, and $\sigma_{{\rm Fe}5015}$.

Errors in calibration to the Lick/IDS system were determined from the
residuals of the standard star offset fits.  As is typical of Lick index 
studies, these calibration errors are significant and often 
dominate the error budget.  However, they are less important for our 
purposes, since we are primarily interested in differential effects.

Although not a concern for the analysis technique of \S~3.1, uncertainties 
within the
Lick/IDS data set are necessary for the fitting procedure of \S~3.2.
Errors for each index were taken from \citet{worthey94} and \citet{wo97}.
Since each standard star was given equal weight in the calibration to the
Lick/IDS system, the IDS internal error was calculated as the average of
the errors in the individual stars, weighted by the number of IDS observations.
As per \citet{worthey94}, errors for IDS standard stars, of which two were 
observed in January and one in April, were reduced by a further factor of
$1.2$.

Errors from all sources are listed individually in Table~2.  From these,
a total error was calculated for each index in each aperture.  For this
purpose the Poisson noise error was ignored as it is already incorporated
within many of the other error sources.  The uncertainty within the Lick/IDS
system, $\sigma_{\rm IDS}$, was also excluded from the formal error quoted
in Table~1 as it represents a purely systematic error.  The remaining sources 
of error were added in quadrature to determine the total observational error.  
Note that although
several of these error sources (e.g. Lick/IDS calibration, sky subtraction,
wavelength calibration) involve systematic corrections, the errors are
random.  For example, sky subtraction involves subtracting the same sky 
spectrum from each aperture on an image.  But it is entirely reasonable to 
suppose that the actual sky intensity will vary slightly from pixel to pixel 
in the spatial direction within each aperture.  Thus the sky at each row of 
the CCD will be over- or undersubtracted by a small and random amount, 
resulting in random sky subtraction error.

\section{Results}

Before proceeding to measure stellar population parameters, gradients in
individual indices can be measured.  The Mg$_2$ index is most commonly used
for this purpose \citep[e.~g.][]{davies93, ka99}.  The Mg$_2$ gradient in
NGC~3610 is presented in Fig.~3; the gradient of 
$\Delta{\rm Mg_2}/\Delta{\rm log}(r/r_e) = -0.056 \pm 0.004$ is in excellent 
agreement with previous work on large samples of elliptical galaxies.  It
is also worth noting that within the off-axis apertures the Mg$_2$ 
measurements are systematically larger in the apertures nearest the major
axis (A2--C2).  

\subsection{Single Stellar Population Models}

The age and mean metallicity of NGC~3610 were estimated from single stellar 
population (SSP) model grids
of \citet*{tmb03} in the H$\beta$--[MgFe]$'$ plane.  Balmer indices such as
H$\beta$ are primarily age-sensitive, while the composite index [MgFe]$'$
was defined by \citet{tmb03} as a metallicity-sensitive index which is not
affected by variations in $[\alpha/{\rm Fe}]$.  Thus by plotting data and
model grids in Balmer index vs. [MgFe]$'$ space (any metal index can be used 
if $[\alpha/{\rm Fe}]$ is not expected to be variable) one can interpolate the
SSP age and metallicity for each observation.
Note that most galaxies did not form in a single starburst, so the SSP 
parameters reflect luminosity-weighted mean values not the true age,
metallicity, and $[\alpha/{\rm Fe}]$ of the galaxy. 
Fig.~4 shows a mean age of approximately 
1.9~Gyr, with ${\rm [Z/H]} \sim +0.2$ averaged over the off-axis data points.
Moreover, there is a clear and significant
gradient within the off-axis apertures in the sense that the stars in the 
outermost apertures are younger and more metal-rich than those in the 
innermost apertures.  This trend is not statistically significant in 
similar grids using H$\gamma$ and H$\delta$ (Figs.~5~\&~6), however.
Note that although the gradients in individual metal lines and the higher
order Balmer lines (see Table~1)
are only marginally statistically significant at best, the H$\beta$
gradient is highly significant.  The average H$\beta$ measurement in apertures
C1 and C2 is higher by more than four standard deviations than the average
H$\beta$ measurement in apertures A1 and A2.

Variations in $[\alpha/{\rm Fe}]$ with radius were measured from the
same models. The 2.0~Gyr models were linearly interpolated in
the Mg~$b$--$<{\rm Fe}>$ plane (Fig.~7).  The offsets perpendicular to the
isoenhancement lines between the 2.0~Gyr models and the 1.0 and 3.0~Gyr
models were used to interpolate an age correction to the final 
$[\alpha/{\rm Fe}]$ measurements.  Five apertures have nearly identical
$[\alpha/{\rm Fe}]\sim0.25$, while aperture C2 is more $\alpha$-enhanced.

The on-axis data set extends the radial range of stellar population
measurements inwards to the center of NGC~3610.  The central $r_e/8$
aperture (labelled Center in Table~1; small square in Figs.~4--7) is in good 
agreement with previous work \citep{hg04, denicolo}.
Metallicity is much higher in the center of the galaxy than in the 
$0.75$--$1.25~r_e$ region, while age and $[\alpha/{\rm Fe}]$ are
found to be consistent with the average values in the outer apertures.  
The two flanking apertures (N and P in Table~1; averaged together and
plotted as the large square in Figs.~4--7) provide an intermediate
measurement between the center and the off-axis apertures.  Metallicity
is between that of the center and the off-axis apertures, but 
$[\alpha/{\rm Fe}]$ is slightly higher than in either the center or the 
off-axis apertures.  Age is unclear --- H$\beta$ and H$\gamma$ suggest a
slightly younger age than the center and most or all of the off-axis
apertures, while H$\delta$ suggests that all three regions are coeval.
On average the three regions of NGC~3610 ($r_e/8$, intermediate,
and off-axis apertures) show a metallicity gradient with no significant
trends in age or $[\alpha/{\rm Fe}]$.  However the detailed radial stellar 
population profile, taking into account the individual off-axis apertures
apparently becoming younger and more metal rich with radius (Fig.~4), is 
implausibly variable.  Metallicity decreases outwards to the off-axis 
apertures, then increases; age decreases slightly, then increases at 
the innermost off-axis apertures and sharply decreases.

It is important to rule out uncorrected instrumental bias
or systematic error as the cause of these unusual gradients.
This possibility has been carefully examined.  Since many age and
metallicity indices, including H$\beta$, Mg$b$, Fe5270, and Fe5335, increase 
with radius one might imagine that the gradients in age and metallicity arise
from oversubtracting sky light.  This was tested
by rereducing the data with 10\% less sky subtracted, an amount far in
excess of the Poisson error in the sky intensity.  Even in this extreme
case, the sense of the gradients measured in this study was preserved, though
naturally the quantitative details were different.  Additional experiments
showed that the gradients are minimized if approximately 20\% less sky
is subtracted.  Furthermore, if sky
oversubtraction was a problem one would expect that {\it all} indices would 
follow the trend of increasing monotonically with radius.  Table~1 shows
that this is not the case.  An incorrect sky
measurement cannot be the source of the observed gradients.

Scattered light is another potential source of instrumental contamination
of our results.  Tests were performed by Javier Cenarro in which stellar
spectra were convolved with the spatial brightness profile of the off-axis 
NGC~3610 data and then extracted and measured identically to the galaxy
apertures (Cenarro, private communication).  Slight gradients due to 
scattered light were detected, but of very small degree, in all cases 
less than the index errors.  These scattered light gradients also generally 
had the opposite sign of the observed galaxy gradients.  We conclude that 
scattered light contamination is not a problem.

Having ruled out the plausible sources of systematic error, two possibilities
remain.  Either NGC~3610 really does have an implausible radially
varying stellar population gradient as suggested by Fig.~4, or the standard
stellar population measurement technique using the H$\beta$, Mg$b$, Fe5270,
and Fe5335 indices is flawed.  Stellar population measurements using H$\gamma$
and H$\delta$ instead of H$\beta$ do not resolve this question.  As shown in
Figs.~5 and 6 respectively, the observations hint at the age gradient within
the off-axis apertures seen in Fig.~4 but not at a statistically significant
level.  An independent stellar population fitting method
allows these possibilities to be addressed.

\subsection{Multiple Index Fitting}

The technique of \citet{proctor04} provides 
a robust method of measuring age, metallicity, and $\alpha$-enhancement
in galaxy spectra.  Briefly, this
involves the creation of a three-dimensional grid of log~$t$, [Fe/H], and
$[\alpha/{\rm Fe}]$, with model values for each Lick index at each point in 
the grid calculated from the solar abundance ratio SSPs of \citet{tmb03} 
and modified according to \citet{tb95} to reflect variations in 
$[\alpha/{\rm Fe}]$.  These $[\alpha/{\rm Fe}]$ calculations were performed
using the Fe- method from \citet{proctor02}.  A ${\chi}^2$ fit is then 
performed to find the best fit model parameters to the set of indices 
observed in a particular galaxy.  The use of every observed index in the fit 
has the advantage of allowing deviant indices, whether from sky subtraction 
residuals, emission contamination, or other problems, to be omitted 
from the fit while still producing reliable measurements of the galaxy 
parameters.  It has been shown that ages can be measured even if all Balmer 
indices are excluded \citep{proctor02}, for example.  For this calculation
the systematic uncertainty within the Lick/IDS system was added in 
quadrature to the observational errors.

No significant gradient is measured in any parameter between 
$0.75$--$1.25~r_e$ (Fig.~8).  Slight offsets (1--2 standard deviations) in 
age and metallicity are 
found between the apertures closest to the galaxy major axis (A2--C2) and 
their counterparts closer to the minor axis, with the major axis apertures
being younger and more metal-rich.  It is possible
the disk population may preferentially add young, metal-rich stars to 
these apertures.  However, the observed disk component of NGC~3610 is
very small \citep{sb90, wmsf97, rw92}, extending to only a few arcseconds.
It is unclear how such a disk could produce much effect at distances beyond 
$1~r_e$.  A gradient in metallicity is found between the center of the galaxy 
and the off-axis apertures, $-0.30 \pm 0.05$ dex in [Z/H] per decade in 
radius.  The gradients in age and $[\alpha/{\rm Fe}]$ have a statistical 
significance of $<2\sigma$ over the radial range,
hinting that the galaxy center might be slightly younger and more 
$\alpha$-enhanced relative to the outer parts.  Age and metallicity were
also found to be quantitatively different than those measured in \S~3.1
in the sense that the multi-index fit measures older ages and lower [Z/H].
H$\beta$ was found to be approximately $0.15~\mbox{\AA}$
too large compared to the best fit value in all apertures.  However,
excluding it or any other index does not alter these 
conclusions but only reduces the ${\chi}^2$ value of the fits.  

As a consistency check, this multi-index fitting procedure was run using
just the H$\beta$, Mg$b$, Fe5270, and Fe5335 indices.  The results were
very similar to those presented in \S~3.1 based on two dimensional grids
of combinations of those four indices (Fig.~9).  The multi-index fits were 
in good quantitative agreement in both age and metallicity, systematically 
offset by $\sim 0.03$~dex in the former and $\sim-0.1$~dex in the latter
relative to the two dimensional fits.  The metallicity gradient 
using just these four indices remained
the same using either fitting technique.  This shows that the different
stellar population measurements from the full multi-index fit are not
artifacts of the fitting software, but reflect real physical information
gained by using the additional ten indices in the fit.

\section {Discussion and Conclusions}

An extremely high $S/N$ long slit spectrum of the outer parts of NGC~3610 has
been obtained using the Keck telescope.  Lick indices have been measured
in six apertures along the slit, spanning the major and minor axes of
the galaxy.  Indices were also measured in three supplementary apertures
on a central spectrum lying along the galaxy major axis.  Stellar population
parameters were measured using both the traditional methods involving two
dimensional grids of a few Lick indices and using a multi-index fitting
procedure that determines age, [Z/H], and $[\alpha/{\rm Fe}]$ simultaneously
using every available index.  Using H$\beta$ as the age-sensitive index,
the traditional method indicated a 
variable and physically implausible radial gradient in age and metallicity,
though only the metallicity gradient was statistically significant if 
higher order Balmer lines were used instead of H$\beta$.
The multi-index fitting technique produced consistent results using
the same limited set of indices, but yielded a different set of
gradients once all available indices were utilized.  The
outer apertures are consistent with a flat population gradient, and when
combined with the interior data points the overall gradients from the
galaxy center out beyond $1~r_e$ are qualitatively consistent with the 
conclusions of previous studies \citep{davies93, ka99, saglia, mehlert03} 
based on samples of other elliptical galaxies: age and $[\alpha/{\rm Fe}]$
do not vary significantly with radius, while [Z/H] decreases with radius.  
The overall gradient in 
[Z/H] is $-0.30$~dex per decade in $r/r_e$, in agreement with the
sample of \citet{ka99} but slightly larger than the typical
values quoted in the other studies.  For example, high redshift formation
models by \citet{pipino} predict shallow gradients, $\sim -0.12$ in cases 
where the $[\alpha/{\rm Fe}]$ gradient is small, as is the case in NGC~3610.

Since mergers are expected to flatten a galaxy's metallicity gradient
\citep{white80}, it is intriguing that NGC~3610 has a relatively steep
gradient despite a wealth of merger remnant signatures (young age, young
globular clusters, much fine structure, stellar disk).  Models of metallicity
gradients in remnants of mergers between pure disk progenitor galaxies 
predict very shallow gradients, $-0.03$ to $-0.13$~dex per decade in 
radius, depending on galaxy mass 
\citep[$10^{10}$--$10^{12} {\rm M}_{\odot}$;][]{bekki}.  The old metal-rich 
globular cluster population of NGC~3610 indicates that at least one progenitor
galaxy had a significant bulge component \citep{strader04}, assuming 
NGC~3610 did indeed form via a major merger event.  However, the presence 
of a bulge might act to inhibit gas flow to the center of the merger, so 
it is not clear that interaction models with bulges would produce steeper
gradients.  The CDM chemodynamical simulations of \citet{kobayashi04} 
indicate that the gradient observed in NGC~3610 is still consistent with
a major merger history, though steeper than typical for major merger
remnants.

A very important ancillary conclusion is that the use of only a few Lick 
indices at a time can produce misleading results, particularly if the
age-sensitive index chosen is H$\beta$.  Several factors
can lead to such biases.  The Balmer lines, traditionally the primary age
indicators among the set of Lick indices, can be contaminated by emission
from star forming regions.  The H$\beta$ index, which in many studies is 
the only Balmer line available, is most strongly affected by such emission.  
As described above, emission corrections can be made
but such corrections are notoriously unreliable.  As shown by Fig.~4, 
this is not a major problem in NGC~3610 as only a few apertures
showed any [OIII]$\lambda{5007}$ emission, and such emission was always
small.  A more serious problem with the H$\beta$ index was pointed out
by \citet{pca}.  Their principal components analysis of Galactic globular
cluster indices suggests a nonlinear component to the relation between
H$\beta$ and metallicity, driven by weak Fe~I lines within the H$\beta$
index passbands.  This nonlinear relation is expected to produce spurious
spreads in ages derived using H$\beta$ compared to ages derived
from higher order Balmer lines --- precisely the effect seen in the off-axis
apertures (Figs. 4--6).  Recall that of the off-axis indices used in
\S3.1, H$\beta$ was the most variable, the only index with radial
variations that were clearly statistically significant.
A study of Galactic globular clusters \citep{pfb04}
has also shown a spurious spread in derived ages resulting from the use of
H$\beta$.

Depending on the set of Lick index standard stars observed,
differences in calibration uncertainty can also yield unreliable results
if poorly calibrated indices are used.  Two indices, H$\delta_{\rm F}$ and 
Fe5015, suffered from this effect in the
case of NGC~3610.  As long as the errors are accurately determined, the
multi-index fitting procedure can use even quite uncertain indices.
Attempting to measure anything using two dimensional grids of such indices
will at best lead to results of low statistical significance.  In principle
several other issues such as night sky line residuals can also contaminate
individual index measurements, and thus any stellar population measurements
made using primarily that index.  Such indices would be easily 
flagged as outliers in a multi-index fitting technique which forces a
stellar population fit to be consistent with every index measurement.

\section{Acknowledgments}\label{sec_ack}

We thank Ricardo Schiavon, Raja Guhathakurta, Glenda Denicol\'{o}, 
Bill Mathews, and Mike Beasley for helpful discussions.  We are grateful to 
Javier Cenarro for running scattered light tests.  DF and RP thank the
ARC for their financial support.  We acknowledge support by the National
Science Foundation through Grant AST-0206139.  We thank the anonymous
referee for helpful comments.

\input epsf.tex
\pagestyle{empty}
\begin{deluxetable}{llllllllllllllll}
\tabletypesize{\footnotesize}
\rotate
\tablecolumns{16}
\tablewidth{0pt}
\tablenum{1}
\tablecaption{Index table}
\tablehead{
\colhead{Aperture} & \colhead{Ca4227}  & \colhead{G4300} & \colhead{Fe4383} & \colhead{Ca4455} & \colhead{Fe4531} & \colhead{C4668} & \colhead{H$\beta$} & \colhead{Fe5015} & \colhead{Mg$_2$} & \colhead{Mg$b$} & \colhead{Fe5270} & \colhead{Fe5335} & \colhead{Fe5406} & \colhead{H$\delta_F$} & \colhead{H$\gamma_F$} \\
\colhead{Errors} & \\
          }
\startdata
A1 & 0.81 & 4.68 & 4.46 & 1.14 & 3.12 & 4.52 & 2.38 & 4.94 & 0.205 & 3.27 & 2.55 & 2.00 & 1.31 & 1.62 & -0.04 \\
$\sigma$ & 0.05 & 0.10 & 0.12 & 0.10 & 0.05 & 0.08 & 0.04 & 0.14 & 0.002 & 0.07 & 0.05 & 0.09 & 0.06 & 0.21 & 0.03 \\
A2 & 0.80 & 4.64 & 4.41 & 1.17 & 3.11 & 4.70 & 2.40 & 4.77 & 0.209 & 3.36 & 2.64 & 2.09 & 1.36 & 1.65 & -0.01 \\
$\sigma$ & 0.05 & 0.10 & 0.12 & 0.10 & 0.05 & 0.08 & 0.04 & 0.14 & 0.002 & 0.07 & 0.05 & 0.09 & 0.06 & 0.21 & 0.03 \\
B1 & 0.87 & 4.68 & 4.39 & 1.15 & 3.02 & 4.57 & 2.35 & 4.84 & 0.203 & 3.32 & 2.60 & 2.00 & 1.37 & 1.60 & -0.05 \\
$\sigma$ & 0.05 & 0.10 & 0.12 & 0.10 & 0.05 & 0.08 & 0.04 & 0.14 & 0.002 & 0.07 & 0.05 & 0.09 & 0.06 & 0.21 & 0.03 \\
B2 & 0.75 & 4.75 & 4.42 & 1.20 & 3.06 & 4.79 & 2.47 & 4.72 & 0.209 & 3.36 & 2.64 & 2.07 & 1.35 & 1.75 & 0.00 \\
$\sigma$ & 0.05 & 0.10 & 0.12 & 0.11 & 0.05 & 0.08 & 0.05 & 0.14 & 0.002 & 0.07 & 0.05 & 0.09 & 0.06 & 0.21 & 0.03 \\
C1 & 0.83 & 4.67 & 4.55 & 1.22 & 3.13 & 4.33 & 2.55 & 4.78 & 0.196 & 3.36 & 2.74 & 2.04 & 1.37 & 1.80 & 0.00 \\
$\sigma$ & 0.05 & 0.11 & 0.13 & 0.11 & 0.06 & 0.08 & 0.05 & 0.14 & 0.002 & 0.07 & 0.05 & 0.09 & 0.07 & 0.22 & 0.04 \\
C2 & 0.80 & 4.73 & 4.74 & 1.31 & 3.21 & 4.81 & 2.70 & 4.79 & 0.210 & 3.52 & 2.74 & 2.01 & 1.41 & 1.71 & 0.03 \\
$\sigma$ & 0.05 & 0.12 & 0.14 & 0.11 & 0.06 & 0.09 & 0.05 & 0.14 & 0.002 & 0.08 & 0.05 & 0.09 & 0.06 & 0.21 & 0.04 \\
All & 0.76 & 4.81 & 4.25 & 1.44 & 2.94 & 4.60 & 2.54 & 4.91 & 0.206 & 3.36 & 2.63 & 2.13 & 1.37 & 1.58 & -0.06 \\
$\sigma$ & 0.05 & 0.09 & 0.11 & 0.10 & 0.05 & 0.08 & 0.04 & 0.14 & 0.002 & 0.07 & 0.05 & 0.09 & 0.06 & 0.21 & 0.03 \\
Center & 0.86 & 5.57 & 5.14 & 1.78 & 3.92 & 7.46 & 2.38 & 6.26 & 0.270 & 4.05 & 2.86 & 2.60 & 1.73 & 1.01 & -0.52 \\
$\sigma$ & 0.09 & 0.23 & 0.09 & 0.11 & 0.19 & 0.21 & 0.03 & 0.27 & 0.004 & 0.14 & 0.18 & 0.07 & 0.04 & 0.19 & 0.05 \\
N & 0.81 & 5.07 & 4.45 & 1.51 & 3.77 & 5.60 & 2.57 & 5.60 & 0.241 & 3.96 & 2.65 & 2.15 & 1.49 & 1.26 & -0.01 \\
$\sigma$ & 0.09 & 0.24 & 0.13 & 0.12 & 0.19 & 0.22 & 0.04 & 0.27 & 0.004 & 0.14 & 0.18 & 0.07 & 0.06 & 0.20 & 0.06 \\
P & 0.82 & 5.18 & 4.53 & 1.58 & 3.75 & 5.84 & 2.51 & 5.86 & 0.236 & 3.80 & 2.64 & 2.14 & 1.46 & 1.35 & 0.06 \\
$\sigma$ & 0.09 & 0.24 & 0.13 & 0.12 & 0.19 & 0.22 & 0.04 & 0.27 & 0.004 & 0.14 & 0.18 & 0.07 & 0.06 & 0.20 & 0.06 \\

\enddata

\end{deluxetable}

\input epsf.tex
\pagestyle{empty}
\begin{deluxetable}{llllllllllllllll}
\tabletypesize{\footnotesize}
\rotate
\tablecolumns{16}
\tablewidth{0pt}
\tablenum{2}
\tablecaption{Error table}
\tablehead{
\colhead{Error} & \colhead{Ca4227}  & \colhead{G4300} & \colhead{Fe4383} & \colhead{Ca4455} & \colhead{Fe4531} & \colhead{C4668} & \colhead{H$\beta$} & \colhead{Fe5015} & \colhead{Mg$_2$} & \colhead{Mg$b$} & \colhead{Fe5270} & \colhead{Fe5335} & \colhead{Fe5406} & \colhead{H$\delta_F$} & \colhead{H$\gamma_F$} \\
          }
\startdata
\multicolumn{15}{c}{Aperture A1: 0.77 $r_e(-33^\circ)$}\\
$\sigma_{\rm pois,3}$ & 0.01 & 0.02 & 0.06 & 0.03 & 0.03 & 0.03 & 0.03 & 0.05 & 0.001 & 0.02 & 0.04 & 0.03 & 0.01 & 0.05 & 0.01 \\
$\sigma_{\rm sky,3}$ & 0.004 & 0.03 & 0.03 & 0.003 & 0.01 & 0.02 & 0.01 & 0.008 & 0.0 & 0.01 & 0.01 & 0.007 & 0.003 & 0.008 & 0.008 \\

\multicolumn{15}{c}{Aperture A2: 0.75 $r_e(-16^\circ)$}\\
$\sigma_{\rm pois,4}$ & 0.02 & 0.04 & 0.05 & 0.03 & 0.03 & 0.04 & 0.02 & 0.06 & 0.002 & 0.01 & 0.02 & 0.02 & 0.04 & 0.06 & 0.02 \\
$\sigma_{\rm sky,4}$ & 0.006 & 0.03 & 0.03 & 0.004 & 0.01 & 0.02 & 0.01 & 0.008 & 0.0 & 0.01 & 0.01 & 0.007 & 0.003 & 0.008 & 0.008 \\

\multicolumn{15}{c}{Aperture B1: 0.91 $r_e(-54^\circ)$}\\
$\sigma_{\rm pois,2}$ & 0.03 & 0.03 & 0.06 & 0.04 & 0.05 & 0.03 & 0.03 & 0.08 & 0.002 & 0.03 & 0.05 & 0.03 & 0.07 & 0.02 & 0.04 \\
$\sigma_{\rm sky,2}$ & 0.007 & 0.03 & 0.04 & 0.004 & 0.01 & 0.02 & 0.01 & 0.01 & 0.0 & 0.02 & 0.01 & 0.007 & 0.004 & 0.01 & 0.01 \\

\multicolumn{15}{c}{Aperture B2: 0.86 $r_e(0^\circ)$}\\
$\sigma_{\rm pois,5}$ & 0.02 & 0.04 & 0.1 & 0.02 & 0.07 & 0.04 & 0.04 & 0.04 & 0.002 & 0.04 & 0.05 & 0.02 & 0.03 & 0.03 & 0.03 \\
$\sigma_{\rm sky,5}$ & 0.007 & 0.04 & 0.04 & 0.006 & 0.02 & 0.02 & 0.02 & 0.01 & 0.0 & 0.02 & 0.02 & 0.01 & 0.004 & 0.01 & 0.01 \\
$\sigma_{\rm emission,5}$ & -- & -- & -- & -- & -- & -- & 0.02 & 0.02 & -- & -- & -- & -- & -- & 0.006 & 0.01 \\

\multicolumn{15}{c}{Aperture C1: 1.26 $r_e(-81^\circ)$}\\
$\sigma_{\rm pois,1}$ & 0.05 & 0.01 & 0.10 & 0.02 & 0.04 & 0.08 & 0.02 & 0.10 & 0.003 & 0.05 & 0.06 & 0.03 & 0.03 & 0.06 & 0.05 \\
$\sigma_{\rm sky,1}$ & 0.01 & 0.06 & 0.07 & 0.007 & 0.03 & 0.02 & 0.03 & 0.03 & 0.0 & 0.01 & 0.008 & 0.02 & 0.02 & 0.04 & 0.03 \\
$\sigma_{\rm emission,1}$ & -- & -- & -- & -- & -- & -- & 0.01 & 0.01 & -- & -- & -- & -- & -- & 0.003 & 0.005 \\

\multicolumn{15}{c}{Aperture C2: 1.23 $r_e(14^\circ)$}\\
$\sigma_{\rm pois,6}$ & 0.04 & 0.03 & 0.05 & 0.03 & 0.06 & 0.07 & 0.07 & 0.10 & 0.003 & 0.03 & 0.06 & 0.05 & 0.07 & 0.04 & 0.02 \\
$\sigma_{\rm sky,6}$ & 0.01 & 0.07 & 0.08 & 0.01 & 0.03 & 0.04 & 0.03 & 0.02 & 0.0 & 0.04 & 0.03 & 0.02 & 0.008 & 0.02 & 0.02 \\
$\sigma_{\rm emission,6}$ & -- & -- & -- & -- & -- & -- & 0.01 & 0.01 & -- & -- & -- & -- & -- & 0.003 & 0.005 \\

\multicolumn{15}{c}{Errors independent of aperture}\\
$\sigma_{\lambda}$ & 0.02 & 0.05 & 0.06 & 0.08 & 0.02 & 0.02 & 0.02 & 0.06 & 0.0 & 0.0 & 0.008 & 0.006 & 0.0 & 0.02 & 0.01 \\
$\sigma_{v_r}$ & 0.02 & 0.04 & 0.05 & 0.06 & 0.02 & 0.02 & 0.01 & 0.04 & 0.0 & 0.0 & 0.006 & 0.005 & 0.0 & 0.02 & 0.01 \\
$\sigma_{\sigma}$ & 0.009 & 0.0 & 0.0 & 0.01 & 0.0 & 0.0 & 0.0 & 0.0 & 0.0 & 0.02 & 0.0 & 0.03 & 0.02 & 0.0 & 0.0 \\
$\sigma_{\rm cal}$ & 0.04 & 0.07 & 0.08 & 0.03 & 0.04 & 0.07 & 0.03 & 0.12 & 0.002 & 0.07 & 0.04 & 0.08 & 0.06 & 0.21 & 0.03 \\
$\sigma_{\rm IDS}$ & 0.08 & 0.12 & 0.17 & 0.08 & 0.13 & 0.20 & 0.07 & 0.14 & 0.003 & 0.07 & 0.09 & 0.08 & 0.06 & 0.13 & 0.10 \\
\enddata

\end{deluxetable}

\vbox{
\begin{center}
\leavevmode
\hbox{%
\epsfxsize=9.2cm
\epsffile{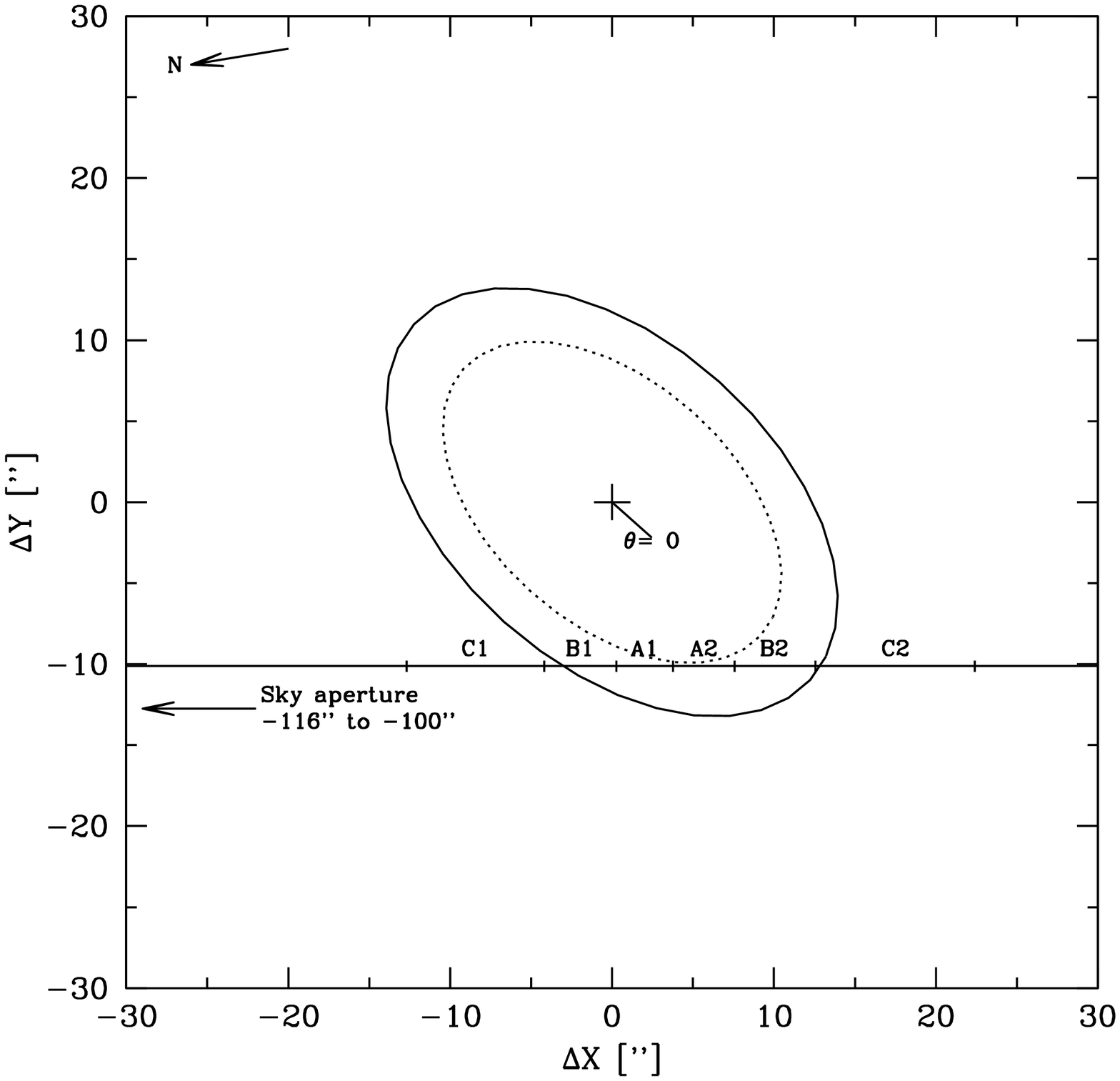}}
\figcaption{\small
A diagram showing the orientation of the off-axis slit from the January data 
with respect to NGC~3610.  The
solid ellipse is a contour of $1~r_e$, and the dotted ellipse is
a contour of $0.75~r_e$.  The horizontal line shows the
position of the slit, with tickmarks at the bounds of each of the six
apertures.  Apertures A1 and A2 are the closest to the galaxy center, on the
minor and major axis sides respectively; the B and C apertures follow the
same numbering pattern at progressively larger galactocentric radii.  The 
angle $\theta$ with respect to the galaxy center is defined relative to the 
southeastern semimajor axis.
}
\end{center}}

\vbox{
\begin{center}
\leavevmode
\hbox{%
\epsfxsize=15cm
\epsffile{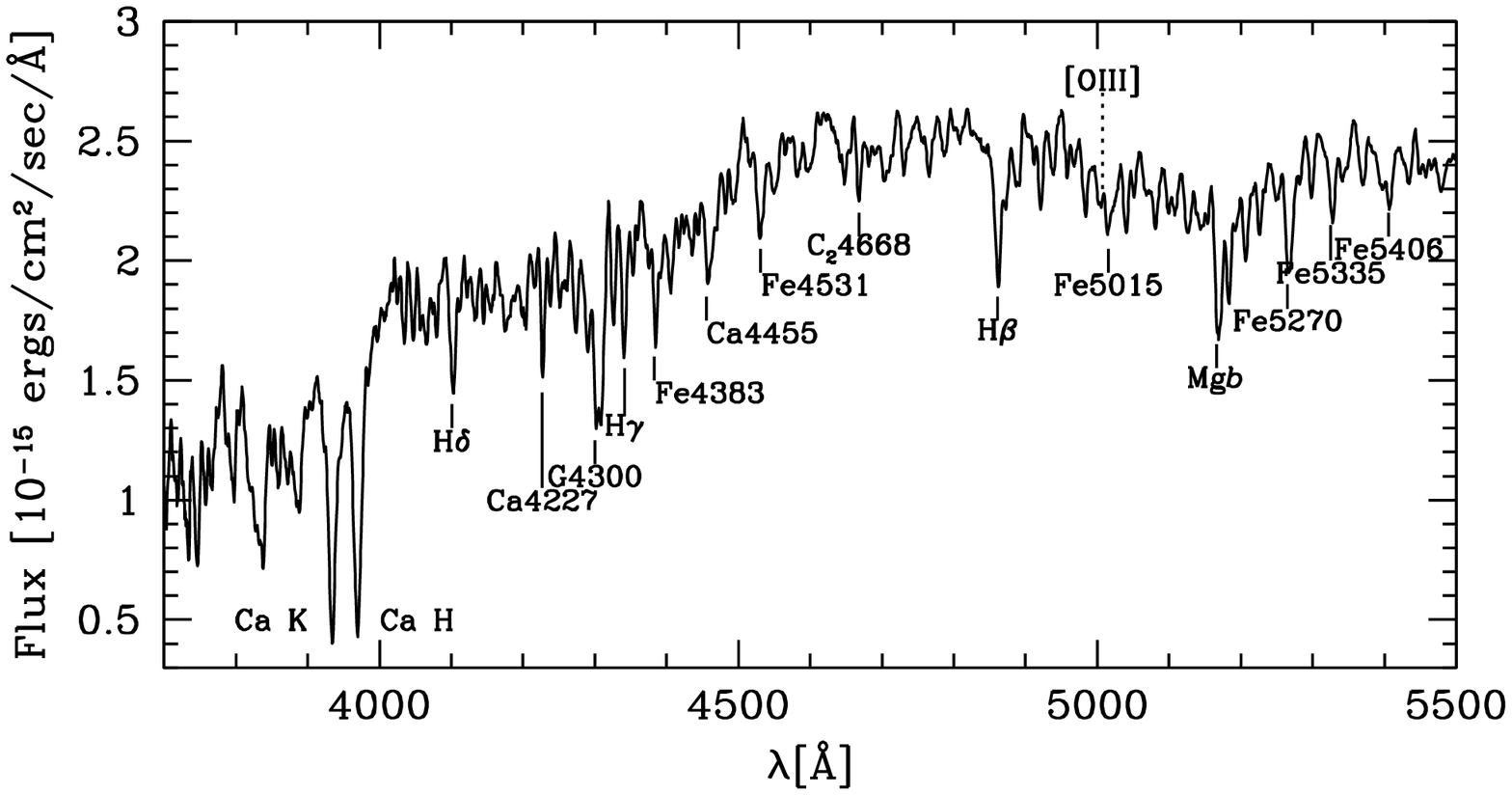}}
\figcaption{\small
Combined spectrum using all off-axis apertures, with the center of each 
spectral index marked.  The
position of the [OIII]$\lambda{5007}$ emission line (of negligible intensity
in this spectrum) is also marked for reference.  $S/N$ per pixel near the
H$\beta$ line is $>700$.
}
\end{center}}

\vbox{
\begin{center}
\leavevmode
\hbox{%
\epsfxsize=9.2cm
\epsffile{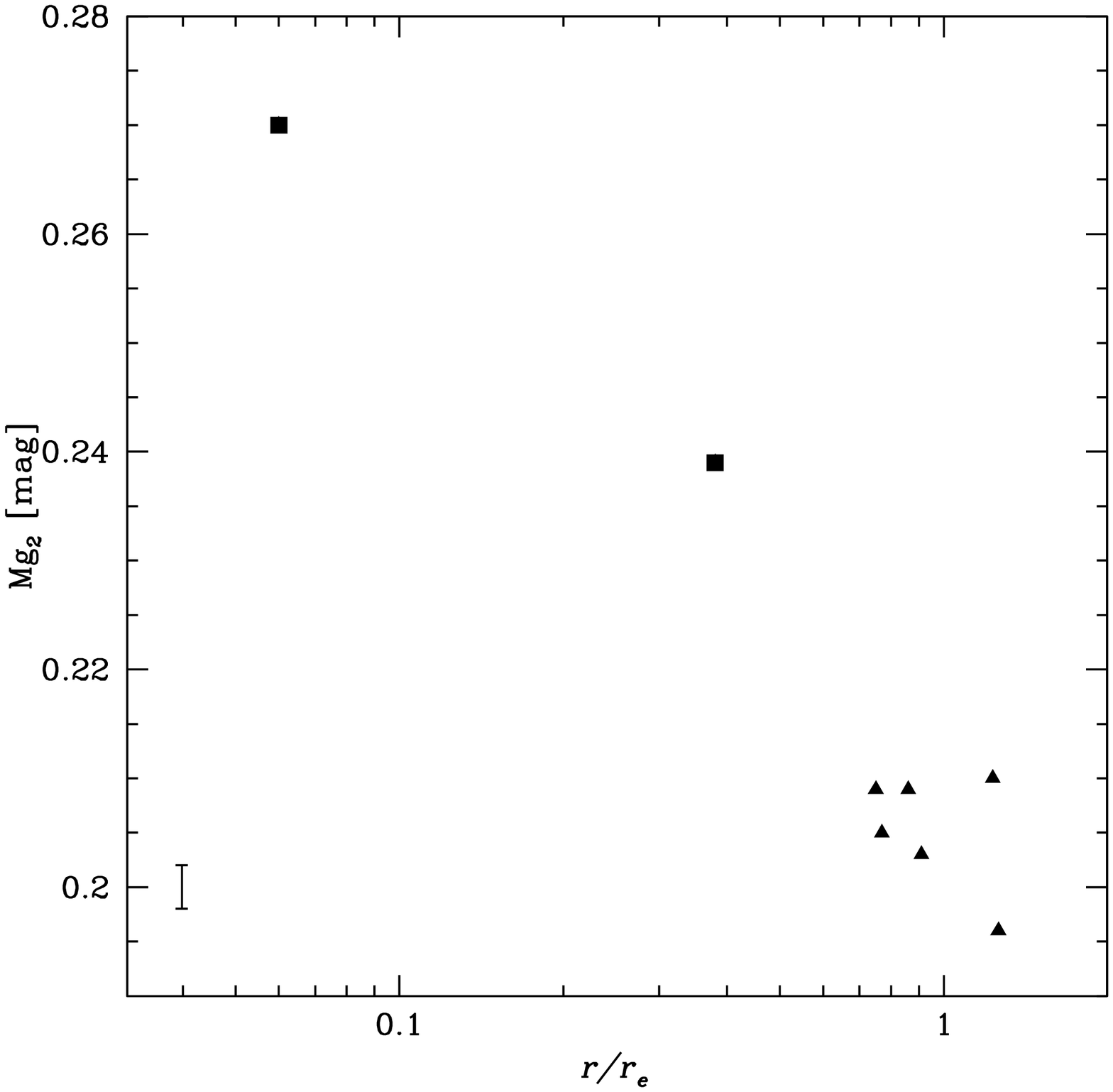}}
\figcaption{\small
Line strength gradient of Mg$_2$.  Data from the on-axis spectrum is 
presented as squares (apertures N and P have been averaged together
and appear as a single point), while data from the off-axis observations
appear as triangles.  Typical errors are presented in the lower left.
A significant radial gradient is visible, $-0.056\pm0.004$~magnitudes
per decade in radius.  This is in excellent agreement with the findings
of \citet{davies93} and \citet{ka99} for Mg$_2$ gradients in large samples
of elliptical galaxies.  Within the off-axis apertures, the apertures
closest to the major axis (A2--C2) have the most Mg$_2$ absorption.
}
\end{center}}

\vbox{
\begin{center}
\leavevmode
\hbox{%
\epsfxsize=9.2cm
\epsffile{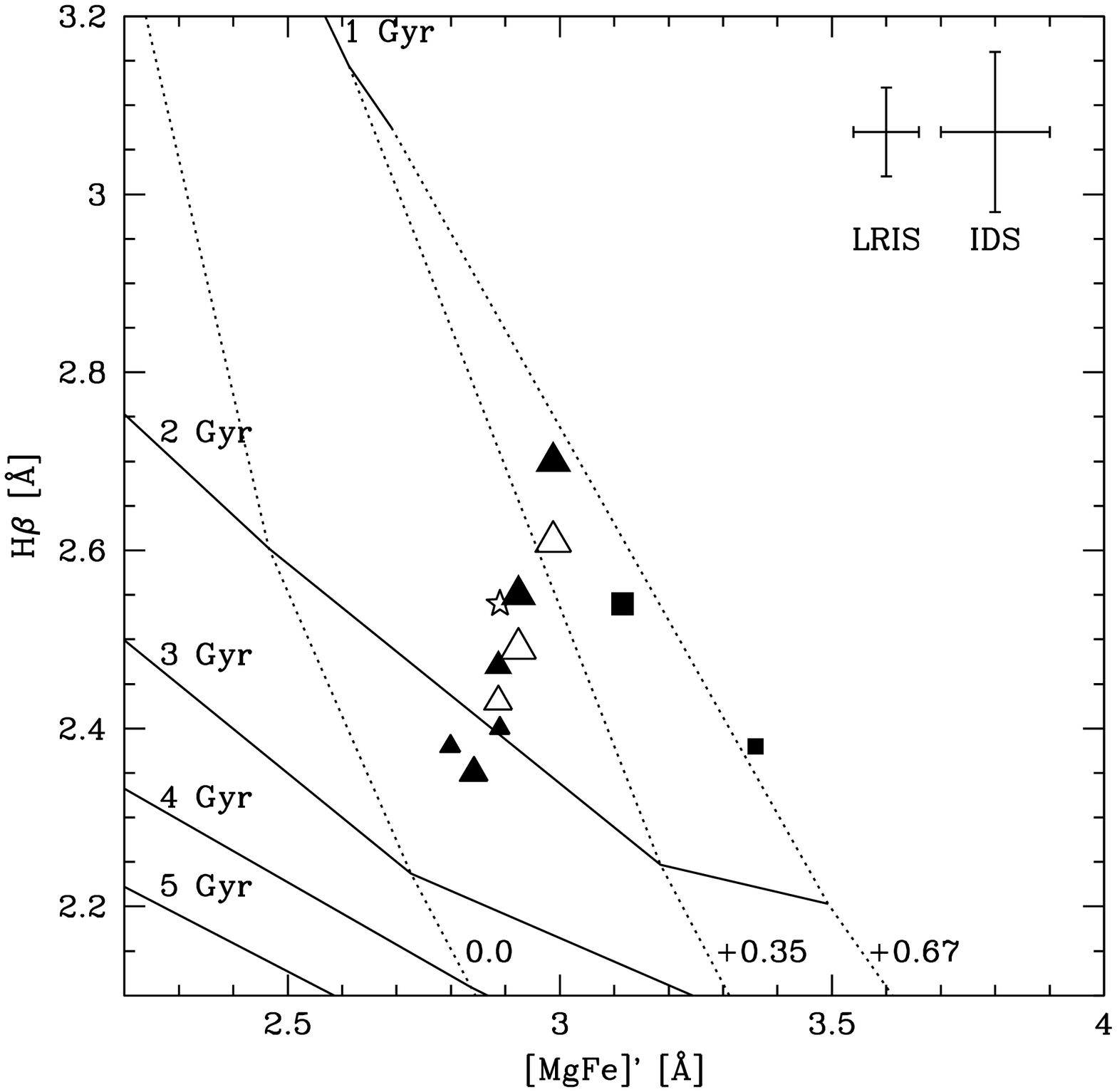}}
\figcaption{\small
H$\beta$ vs. [MgFe]$'$.  LRIS data from the off-axis (triangles) and on-axis
(squares) are presented.
Within each data set, point size is proportional to distance from the
galaxy center.  The open triangles
do not include any H$\beta$ emission correction but are otherwise identical
to the corresponding solid triangles.  The star symbol 
indicates the value of these indices in the combined spectrum shown in Fig.~2.  
Models are from \citet{tmb03}, with solar abundance ratios; isochrone (solid)
and isometallicity (dotted) lines are labeled.  Typical error bars are shown 
in the upper 
right.  The LRIS error bars correspond to the total error in the text, while
the IDS error bars also account for the uncertainty within the Lick/IDS
system; see \S~2.3 for details.
Within the off-axis data set, there appears to be a radial stellar 
population gradient such that the outer apertures are systematically
younger and more metal-rich than the apertures closest to the galactic center.
The central and intermediate radius on-axis apertures are inconsistent
with this picture, being more metal-rich than all but the outermost off-axis
aperture, and of comparable age to the average of the off-axis apertures.
}
\end{center}}

\vbox{
\begin{center}
\leavevmode
\hbox{%
\epsfxsize=9.2cm
\epsffile{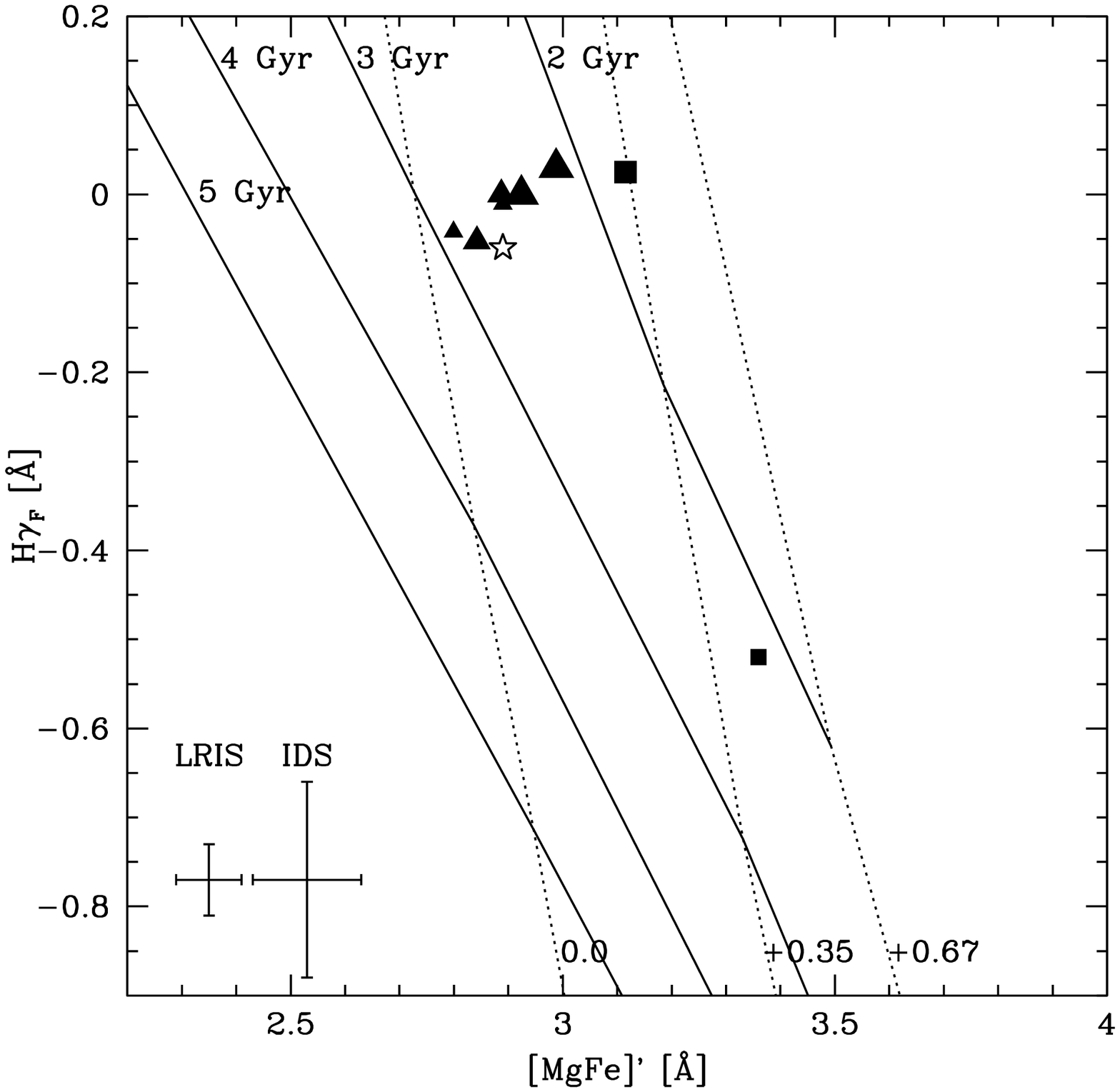}}
\figcaption{\small
H$\gamma_{\rm F}$ vs. [MgFe]$'$.  Points, error bars, and models are as Fig.~3. 
Although qualitatively similar to the trends shown in Fig.~3, the
gradient within the off-axis apertures is not statistically significant in
the H$\gamma_{\rm F}$--[MgFe]$'$ plane.  This figure suggests a monotonic
decrease in metallicity with radius, with only slight variations in age.
}
\end{center}}

\vbox{
\begin{center}
\leavevmode
\hbox{%
\epsfxsize=9.2cm
\epsffile{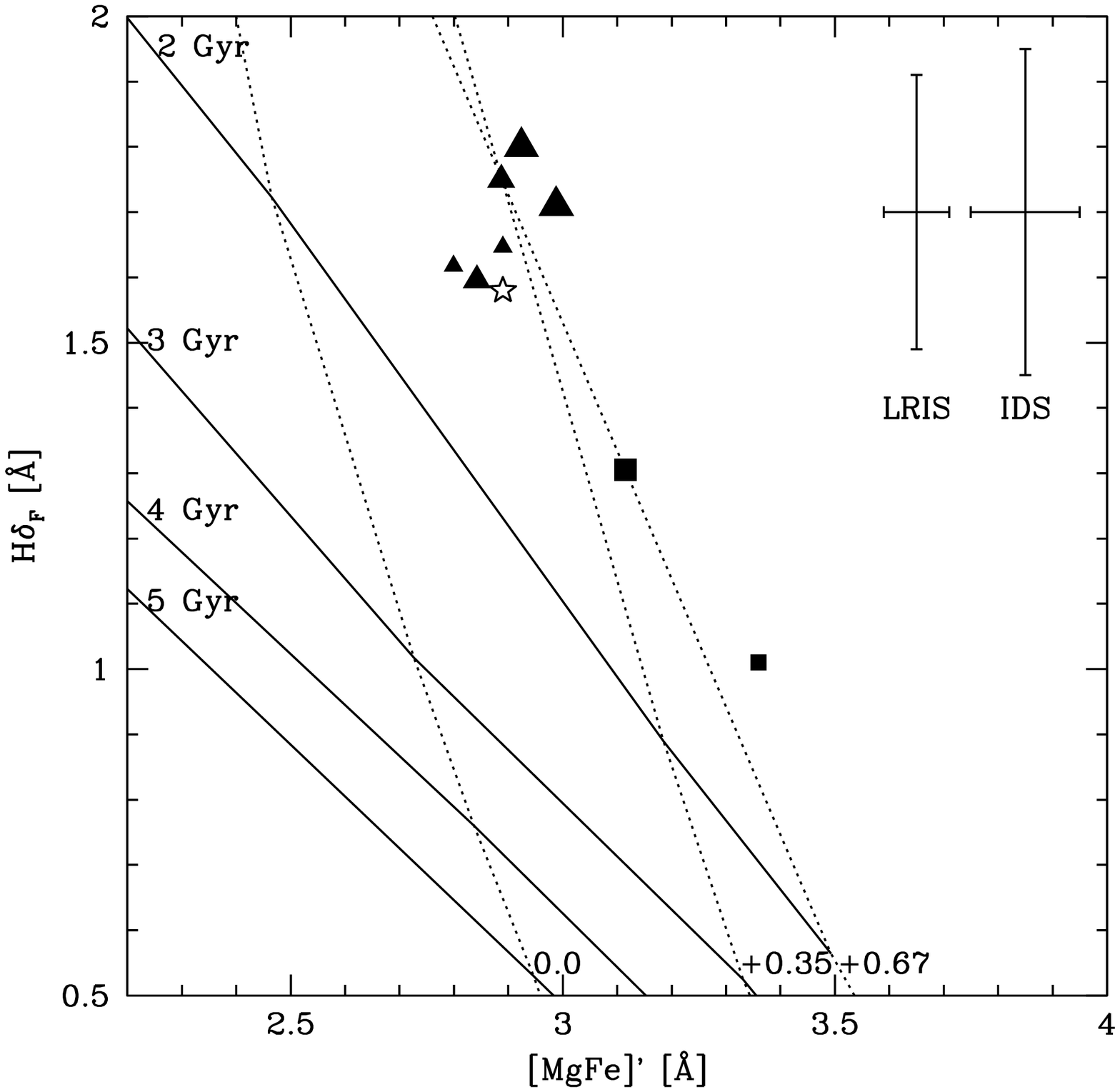}}
\figcaption{\small
H$\delta_{\rm F}$ vs. [MgFe]$'$.  Points, error bars, and models are as Fig~3.  
Within the
errors, the index measurements in the H$\delta_{\rm F}$--[MgFe]$'$ plane are 
consistent with metallicity decreasing with radius, with no significant age 
gradient.
}
\end{center}}

\vbox{
\begin{center}
\leavevmode
\hbox{%
\epsfxsize=9.2cm
\epsffile{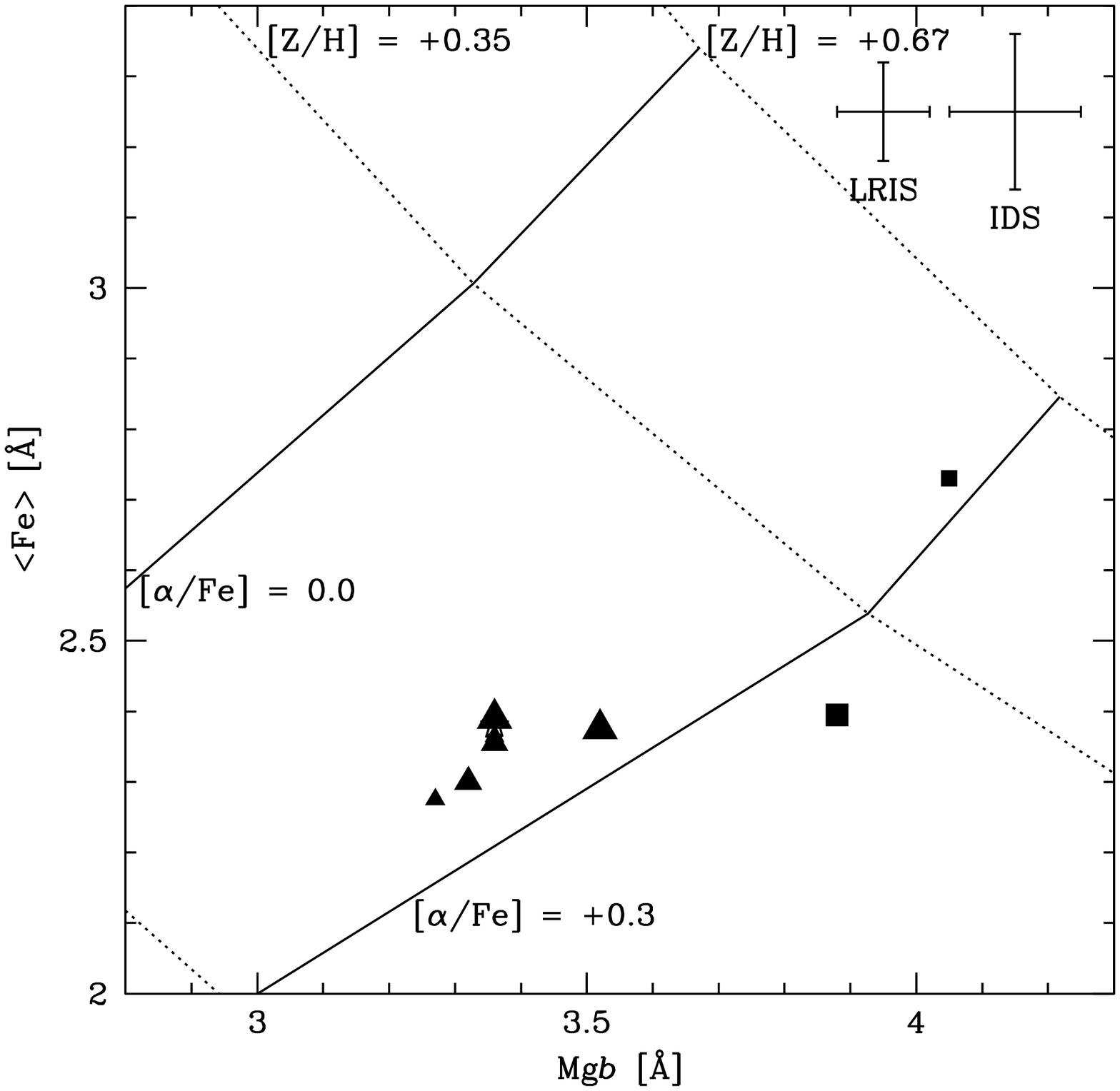}}
\figcaption{\small
$<{\rm Fe}>$ vs. Mg~$b$.  Data points are as Fig~3.  Also shown are 2 Gyr model
grids from \citet{tmb03}.
Solid lines have constant $[\alpha/{\rm Fe}]$, dotted lines have constant [Z/H].
Typical index errors are shown in the top right.  Accounting for the 
differences in age, the off-axis apertures have equivalent $[\alpha/{\rm Fe}]$
with the exception of aperture C2 which is enhanced by $\sim~1$~standard
deviation with respect to the other apertures.  
The $r_e/8$ aperture shows an $[\alpha/{\rm Fe}]$ value consistent with 
the off-axis apertures, while
the intermediate radius apertures yield $[\alpha/{\rm Fe}]$ values 
$\sim~1$~standard deviation higher than either the central or the off-axis 
apertures.
We conclude that there is no significant radial gradient in $[\alpha/{\rm Fe}]$.
}
\end{center}}

\vbox{
\begin{center}
\leavevmode
\hbox{%
\epsfxsize=9.2cm
\epsffile{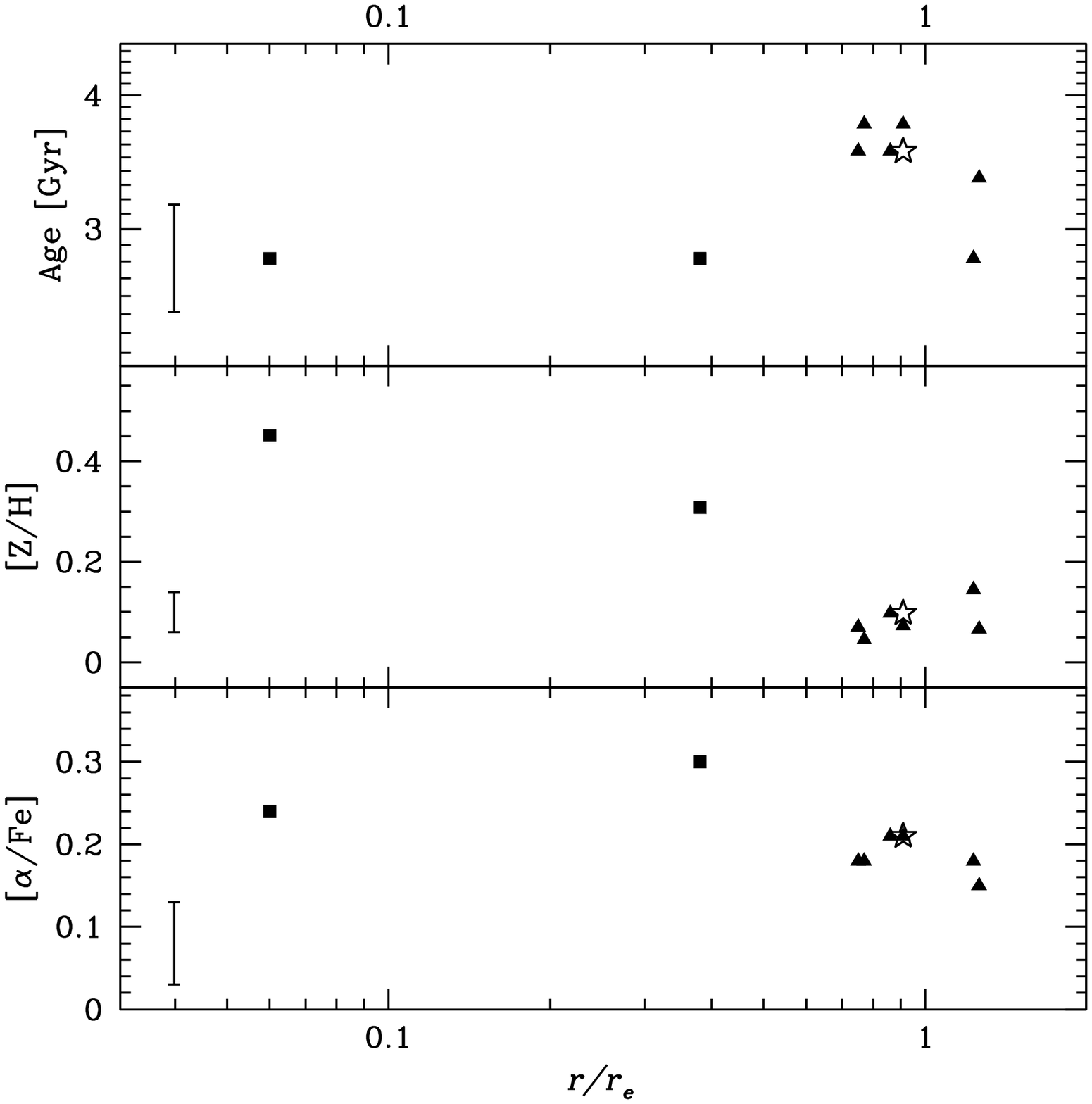}}
\figcaption{\small
Radial gradients in age (top panel), [Z/H] (middle panel), and 
$[\alpha/{\rm Fe}]$ (bottom panel), derived from the multi-index fitting
method, \S~3.2.  Point types are as in Fig.~3.
No significant gradient is detected in age
or $[\alpha/{\rm Fe}]$, though the stellar populations near $0.8$--$0.9~r_e$
are somewhat older than the other apertures.  
A metallicity gradient of $-0.30 \pm 0.05$ dex in 
[Z/H] per decade in radius is detected, though within the off-axis
apertures solely the measurements are consistent with no metallicity gradient.
}
\end{center}}

\vbox{
\begin{center}
\leavevmode
\hbox{%
\epsfxsize=9.2cm
\epsffile{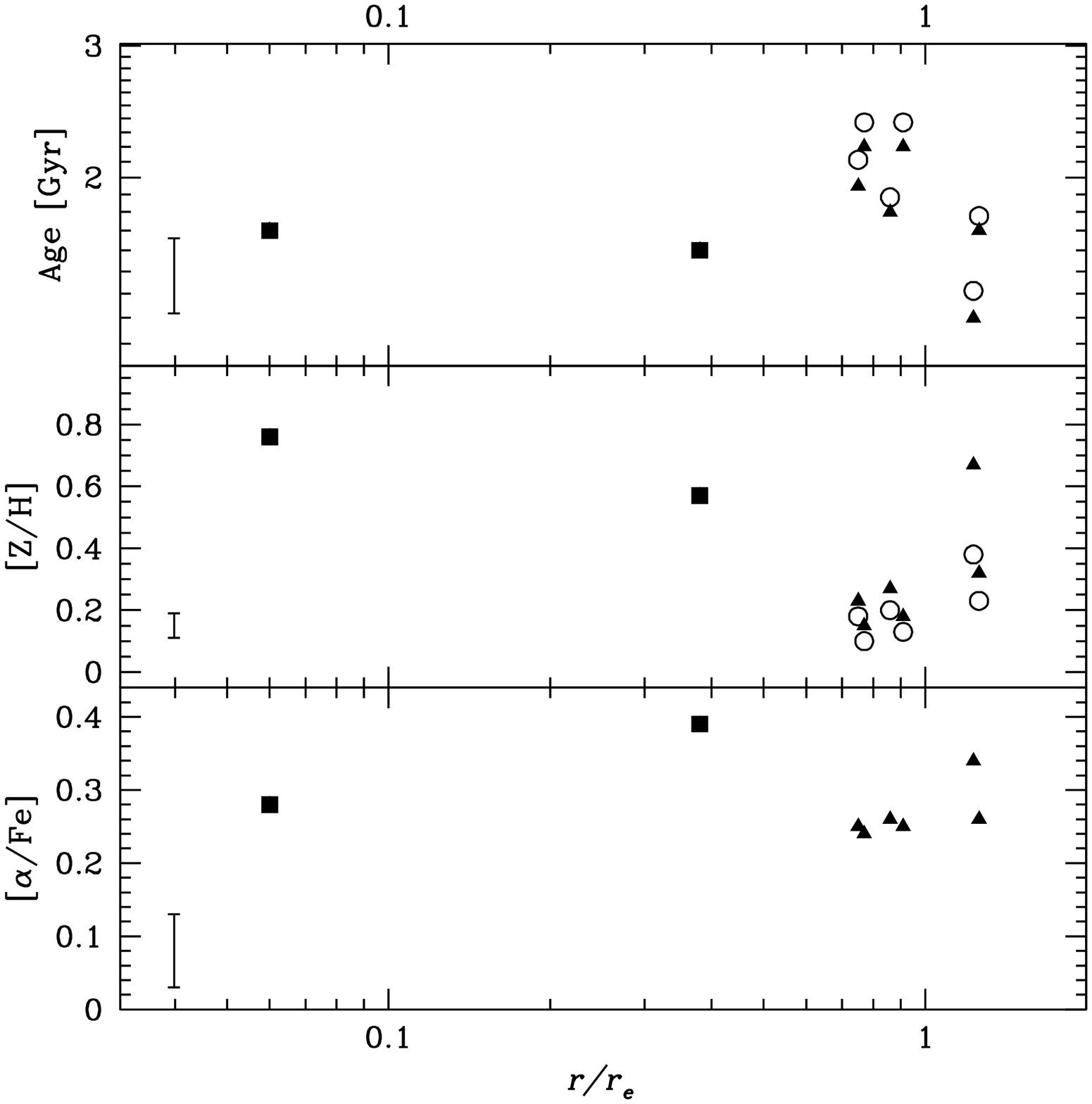}}
\figcaption{\small
Radial gradients in age (top panel), [Z/H] (middle panel), and 
$[\alpha/{\rm Fe}]$ (bottom panel).  Solid point types are as in Fig.~3, 
measured using the two-dimensional fitting technique of \S~3.1 with H$\beta$
as the age-sensitive index.  
The open circle symbols represent the off-axis apertures measured
by applying the multi-index fitting method to only the four indices (H$\beta$,
Mg$b$, Fe5270, and Fe5335) used in the two-dimensional fitting technique.
These fits set $[\alpha/{\rm Fe}]$ to zero since the number of input indices
was inadequate to fit for all three quantities.  Typical error bars in each
quantity are shown in the lower left of each panel.  The two measurement 
techniques differ only by an offset of less than one standard deviation.  The
different results compared to Fig.~8 are entirely due to the more restricted
set of indices used here.  Using just these indices,
the overall age gradient is unclear, though the off-axis apertures show a
steep gradient becoming younger at larger radii.  The overall metallicity
gradient is qualitatively similar to that measured in Fig.~8, but $50\%$
steeper in slope.  In addition, the off-axis apertures show a clear
gradient in the opposite direction whereas Fig.~8 showed that [Z/H]
remains roughly constant in these apertures.  The $[\alpha/{\rm Fe}]$
gradient remains negligible when measured using only this restricted set of 
indices.
}
\end{center}}


\begin{thebibliography}{}
%
\bibitem[Barnes(2002)]{barnes02} Barnes, J.~E.\ 2002, \mnras, 
333, 481
%
\bibitem[Bekki \& Shioya(1999)]{bekki} Bekki, K.~\& Shioya, 
Y.\ 1999, \apj, 513, 108
%
%
\bibitem[Bruzual \& Charlot(2003)]{bc03} Bruzual, G.~\& 
Charlot, S.\ 2003, \mnras, 344, 1000
%
\bibitem[Carlberg(1984)]{carlberg} Carlberg, R.~G.\ 1984, \apj, 
286, 403
%
\bibitem[Carollo et~al.(1993)Carollo, Danziger \& Buson]{carollo93} Carollo, 
C.~M., Danziger, I.~J. \& Buson, L.\ 1993, \mnras, 265, 553
%
\bibitem[Davies et~al.(1993)Davies, Sadler, \& Peletier]{davies93} Davies, 
R.~L., Sadler, E.~M. \& Peletier, R.~F.\ 1993, \mnras, 262, 650
%
\bibitem[Denicol\'{o} et~al.(2004)]{denicolo} Denicol\'{o}, G., Terlevich, R., 
Terlevich, E., Forbes, D.~A., Terlevich, A. \& Carrasco, L. 2004, \mnras,
in preparation
%
\bibitem[Faber et~al.(1989)]{faber89} Faber, S.~M., Wegner, G., Burstein, D.,
Davies, R.~L., Dressler, A., Lynden-Bell, D. \& Terlevich, R.~J. 1989,
\apjs, 69, 763
%
\bibitem[Goudfrooij et al.(1994)]{goud94} Goudfrooij, P., 
Hansen, L., Jorgensen, H.~E., Norgaard-Nielsen, H.~U., de Jong, T. \& van 
den Hoek, L.~B.\ 1994, \aaps, 104, 179 
%
%
\bibitem[Howell \& Guhathakurta(2004)]{hg04} Howell, J.~H. \& 
Guhathakurta,~P. 2004, in preparation
%
\bibitem[Idiart, Michard \& de Freitas Pacheco(2002)]{idiart02} 
Idiart, T.~P., Michard, R. \& de Freitas Pacheco, J.~A.\ 2002, \aap, 383, 
30
%
\bibitem[Jones(1999)]{jones} Jones, L.A. 1999, PhD Thesis, University of
North Carolina
%
\bibitem[Jorgensen(1997)]{jorg97} Jorgensen, I.\ 1997, \mnras, 
288, 161
%
\bibitem[Kobayashi \& Arimoto(1999)]{ka99} Kobayashi, C.~\& 
Arimoto, N.\ 1999, \apj, 527, 573
%
\bibitem[Kobayashi(2004)]{kobayashi04} Kobayashi, C.\ 2004, \mnras, 
347, 740
%
\bibitem[Kuntschner \& Davies(1998)]{kd98} Kuntschner, H.~\& 
Davies, R.~L.\ 1998, \mnras, 295, L29 
%
\bibitem[Kuntschner et al.(2002)]{kuntschner} Kuntschner, H., 
Smith, R.~J., Colless, M., Davies, R.~L., Kaldare, R. \& Vazdekis, A.\ 
2002, \mnras, 337, 172
%
\bibitem[Mehlert et al.(2003)]{mehlert03} Mehlert, D., Thomas, 
D., Saglia, R.~P., Bender, R. \& Wegner, G.\ 2003, \aap, 407, 423
%
\bibitem[Oke et al.(1995)]{lris} Oke, J.~B., et al.\ 1995, 
\pasp, 107, 375
%
%
\bibitem[Pipino \& Matteucci(2004)]{pipino} Pipino, A.~\& 
Matteucci, F.\ 2004, \mnras, 347, 968
%
\bibitem[Proctor et al.(2004)]{proctor04} Proctor, R.~N., Forbes, 
D.~A., Hau, G.~K.~T., Beasley, M.~A., De Silva, G.~M., Contreras, R., \& 
Terlevich, A.~I.\ 2004, \mnras, 349, 1381 
%
\bibitem[Proctor \& Sansom(2002)]{proctor02} Proctor, R.~N.~\& 
Sansom, A.~E.\ 2002, \mnras, 333, 517
%
\bibitem[Proctor, Forbes \& Beasley(2004)]{pfb04} Proctor, R.~N., Forbes,
D.~A., \& Beasley, M.~A. 2004, \mnras, submitted
%
\bibitem[Rich(1998)]{bwid} Rich, R.~M. 1998, ASP Conf.\ Ser.\ 147, Abundance
Profiles: Diagnostic Tools for Galaxy History, ed. D.~Friedli, M.~Edmunds,
C.~Robert \& L.~Drissen (San Francisco: ASP), p.~36
%
\bibitem[Rix \& White(1992)]{rw92} Rix, H.~\& White, 
S.~D.~M.\ 1992, \mnras, 254, 389
%
\bibitem[Saglia et al.(2000)]{saglia} Saglia, R.~P., Maraston, 
C., Greggio, L., Bender, R. \& Ziegler, B.\ 2000, \aap, 360, 911
%
\bibitem[Schiavon (2004)]{ricardo} Schiavon, R. 2004, in preparation
%
\bibitem[Schweizer \& Seitzer(1992)]{ss92} Schweizer, F. \& Seitzer, P. 1992,
\aj, 104, 1039
%
\bibitem[Scorza \& Bender(1990)]{sb90} Scorza, C.~\& Bender, 
R.\ 1990, \aap, 235, 49
%
\bibitem[Silva \& Bothun(1998)]{silbot98} Silva, D.~R.~\& 
Bothun, G.~D.\ 1998, \aj, 116, 2793
%
\bibitem[Strader et~al.(2003)]{strader03} Strader, J., Brodie, J., Schweizer,
F., Larsen, S. \& Seitzer, P. 2003, \aj, 125, 626
%
\bibitem[Strader, Brodie \& Forbes (2004)]{strader04} Strader, J., Brodie, J.
\& Forbes, D.~A. 2004, \aj, 127, 295
%
\bibitem[Strader \& Brodie (2004)]{pca} Strader, J. \& Brodie, J. 2004,
\aj, accepted
%
\bibitem[Terlevich \& Forbes(2002)]{tf02} Terlevich, 
A.~I.~\& Forbes, D.~A.\ 2002, \mnras, 330, 547
%
\bibitem[Thomas et~al.(2003)Thomas, Maraston \& Bender]{tmb03} 
Thomas, D., Maraston, C. \& Bender, R.  2003, \mnras, 339, 897
%
\bibitem[Tonry et~al.(2001)]{tonry01} Tonry, J.~L., Dressler, 
A., Blakeslee, J.~P., Ajhar, E.~A., Fletcher, A.~B., Luppino, G.~A., 
Metzger, M.~R. \& Moore, C.~B.\ 2001, \apj, 546, 681 
%
\bibitem[Toomre \& Toomre(1972)]{toomre72} Toomre, A.~\& Toomre, 
J.\ 1972, \apj, 178, 623
%
\bibitem[Trager et~al.(1998)]{trager98} Trager, S.~C., Worthey, G., Faber, 
S.~M. \& Gonzalez, J.~J. 1998, \apjs, 116, 1
%
\bibitem[Trager et~al.(2000a)]{tfwg1} Trager, S.~C., Faber, S.~M.,
Worthey, G. \& Gonzalez, J.~J. 2000a, \aj, 119, 1645
%
%
\bibitem[Tripicco \& Bell(1995)]{tb95} Tripicco, M.~J.~\& 
Bell, R.~A.\ 1995, \aj, 110, 3035 
%
\bibitem[van der Marel(1994)]{pixfit} van der Marel 1994, MNRAS, 270, 271
%
\bibitem[Vazdekis(1999)]{vazdekis} Vazdekis, A.\ 1999, \apj, 
513, 224
%
%
\bibitem[White(1980)]{white80} White, S.~D.~M.\ 1980, \mnras, 
191, 1P
%
\bibitem[Whitmore et~al.(1997)]{wmsf97} Whitmore, B.~C., Miller, B.~W., 
Schweizer, F. \& Fall, S.~M. 1997, \aj, 114, 1797
%
\bibitem[Worthey, Faber, Gonzalez, \& Burstein(1994)]{worthey94} 
Worthey, G., Faber, S.~M., Gonzalez, J.~J., \& Burstein, D.\ 1994, \apjs, 
94, 687
%
\bibitem[Worthey \& Ottaviani(1997)]{wo97} Worthey, G. \& Ottaviani, D.~L.
1997, \apjs, 111, 377
%
%
\end{thebibliography}
\end{document}